\documentclass[aps,prd,amssymb,cite,
amsfonts,epsf,preprintnumbers,nofootinbib,superscriptaddress]{revtex4}
\usepackage{graphics,epsfig}
\usepackage{graphicx}
\usepackage{dcolumn}
\usepackage{epstopdf}
\usepackage{fixmath}
\usepackage{textcomp}
\newcommand{\be}{\begin{equation}}
\newcommand{\ee}{\end{equation}}
\newcommand{\beq}{\begin{equation}}
\newcommand{\eeq}{\end{equation}}
\newcommand{\bea}{\begin{eqnarray}}
\newcommand{\eea}{\end{eqnarray}}

\def\be{\begin{equation}}
\def\ee{\end{equation}}
\def\ba{\begin{eqnarray}}
\def\ea{\end{eqnarray}}

\begin{document}
\title{  Anti-Evaporation/Evaporation of  $n$-dimensional Reissner-Nordstr\"{o}m Black Hole }

\author{YuHong Fang} 
\email{fangyh23@mail2.sysu.edu.cn}

 \author{Zhiqi Huang} 
\email{huangzhq25@mail.sysu.edu.cn}
\author{HaiTao Miao} 
\email{miaoht3@mail2.sysu.edu.cn}
\author{Naveen K Singh} 
\email{naveen.nkumars@gmail.com}

\affiliation{School of Physics and Astronomy,\\
 Sun Yat-Sen University Zhuhai,  2 Daxue Rd, Tangjia, Zhuhai, China.
  \vspace{15mm}}

\begin{abstract}
  We generalize $f(R)$-theory of anti-evaporation/evaporation for 
a Reissner-Nordstr\"{o}m black hole  in $n$-dimensional space-time. We consider non-linear conformally invariant Maxwell field. By perturbing the fields over Nariai-like space-time associated with degenerate horizon, we describe dynamical behavior of horizon. We show that $f(R)$-gravity can offer both anti-evaporation and evaporation in $n$-dimensional Reissner-Nordstr\"{o}m black hole depending on the dimension $n$ and the functional form of $f(R)$. Furthermore, we argue that, in one class of non-oscillatory solution, stable and unstable anti-evaporation/evaporation exist.  In the other class of oscillatory solution anti-evaporation/evaporation exists only with  instability. The first class of solution may explain a long-lived black hole.
\end{abstract}
\vspace{10mm}

\maketitle

\section{\label{sec:level1}Introduction} One possible candidate which facilitates in exploring the current universe and delving into the early universe is
primordial black hole.  In the early universe, black hole can be formed due to non-linear metric perturbations \cite{Ivanov:1997ia,Bullock:1998mi, Bullock:1996at}, density perturbations \cite{hawking1,hawkingCarr}, the evolution of gravitational bound objects  \cite{Kalashnikov} etc.. The observation of such black holes depends on its mass and anti-evaporation/evaporation properties. Primordial black hole could be a possible component of dark matter \cite{Carr:2016drx,Ivanov:1994pa,Blais:2002nd,Chavda:2002cj}. Primordial black hole  could explain current dark matter density better if one considers a wide range of mass of such black hole not limited only in a particular range. Dark matter may be explained by considering a lower mass range ($10^ 6 M_p$ -$10^{11} M_p$)  \cite{Carr:2016drx}, where $M_p$ is the Planck mass. So far, not much attention was given for primordial black hole with masses smaller than $10^{15} $g in explaining dark matter since they are considered evaporated. The phenomenon of anti-evaporation offers a wider range of mass, since primordial black hole with lower masses may exist in the current universe. The black holes with low masses may survive and 
hence contribute to dark matter density today. Contrary to that, evaporation reduces the chance of the presence of primordial black hole in the current universe and hence its contribution to dark matter density. However, whether anti-evaporation exists is debatable. \\

The existence  of evaporation of a black hole was first proposed by Hawking\cite{Hawking:1974sw}. Later on, in contrast to that, Hawking and Bousso introduced anti-evaporation  which is due to quantum correction \cite{Bousso97} and appears for Nariai space-time \cite{Nariai1,Nariai2} where cosmological horizon and event horizon coincide. In the evaporation process, the black hole reduces its horizon size by emitting radiation through the quantum effect. The phenomenon of anti-evaporation, as its name, has the the properties reverse to that of evaporation \cite{Nojiri:1998ph,Buric:2000cj}.  Grand unified theory is also a theory which explains anti-evaporation\cite{Elizalde:1999dw,Bytsenko:1998md} . However, we will follow $f(R)$ theory in higher dimension for a black hole with multi-horizons, in particular, where these horizons become degenerate.\\

In Ref. \cite{Bousso97,Niemeyer:2000nq,Nojiri:1998ue}, anti-evaporation due to quantum correction is studied by
considering two dimensional one-loop effective action. Here, the calculation is done in s-wave approximation. In addition to that, the appearance of conformal anomaly in four dimension may provide anti-evaporation \cite{Nojiri:1998ph,Nojiri:1998ue,Nojiri:2000ja}. However, anti-evaporation is also possible at classical level in $f(R)$-gravity \cite{Nojiri13,Nojiri:2014jqa,Addaz,Oikonomou:2015lgy}. $f(R)$-gravity may prevent a primordial black hole to be evaporated and assist to be long-lived even having small masses. Some recent efforts are made in Refs. \cite{taishi,Singh:2017qur,Addazi:2017puj}. Despite the fact that anti-evaporation is associated  with instability at classical level in  some theories, e.g. in Ref. \cite{Nojiri:2014jqa}, attentions should  be given at classical level besides at quantum level to search a stable solution. In this paper,
we generalize the possibility of anti-evaporation/evaporation in $f(R)$-gravity in $n$-dimension at classical level.\\

Kaluza-Klein theory was the first theory, where higher dimension was introduced first \cite{Kaluza:1921tu, Klein:1926fj,Appelquist:1987nr} to unify gravity and electromagnetism. Later on, the  idea of higher dimensions became a platform in supergravity \cite{Duff:1986hr} and   string theory \cite{Green:2012oqa,Green:2012pqa} in constructing a unified theory of gravity and other fundamental forces. Inspiring from these  higher dimensional theories,
a lot of progress towards black hole physics have been made\cite{Emparan:2008eg,Reall:2015esa}. Some of the interesting results were found if one studies black hole physics in higher dimensions. For example, there is some possibility of creation of mini higher dimensional black hole at LHC \cite{Kanti:2008eq}.  String theory can  calculate the black hole entropy statistically \cite{Strominger:1996sh}. Furthermore, Schwarzschild, Reissner-Nordstr\"{o}m and Kerr solutions were found in higher dimension \cite{Myers:1986un}. 
Higher dimension was later considered in charged black-hole \cite{Xu:1988ju},  charged black hole in (A)DS spaces \cite{Liu:2003px},  Banados-Teitelboim-Zaneli black hole \cite{Ghosh:2011tt,Hendi:2010px}, radiating black hole \cite{Ghosh:2008zza} etc.. In this paper, we explore the possibility of  evaporation and anti-evaporation of black hole in $f(R)$-gravity in higher dimensions. We generalize the $f(R)$-theory of anti-evaporation for Reissner Nordstr\"{o}m\cite{Nojiri:2014jqa}.  To work with exact analytical solution, we consider conformally invariant Maxwell action in $n$-dimension which constrains the dimension $n$  \cite{Sheykhi:2012zz}. The electric field is obtained in this case the same as obtained in the four dimension. \\ 

Sec. (\ref{sec:level2}) briefly discusses the realization of anti-evaporation in $f(R)$-gravity. We discuss analytical solution of Reissner-Nordstr\"{o}m black hole in $f(R)$-gravity in higher dimensions in Sec. (\ref{sec:level3}) and we also mention about solution for extreme black holes. In  Sec. (\ref{sec:level4}), we write modified equations up to the first order of perturbations and obtain solutions in  $n$-dimension. We  explain the anti-evaporation and evaporation for different values of $n$ and other parameters of the theory. Finally, we conclude in  Sec. (\ref{sec:level5}).

 
\section {Anti-Evaporation in $F(R)$ gravity}
\label{sec:level2}
Generalizing the theory of  Bousso and Hawking \cite{Bousso97} for anti-evaporation,  Odintsov and Nojiri showed its possibility even at classical level \cite{Nojiri13}. In this construction, the authors considered $f(R)$-gravity, which is the basic requirement in explaining anti-evaporation. We can obtain a Nariai-like solution, where cosmological horizon and event horizon coincide, and this solution is associated with the solution for  anti-evaporation. The action corresponding to $f(R)$-gravity and matter with gravitational constant $G$ and Ricci scalar $R$ can be written as,

\be
S=\frac{1}{16\pi G}\int d^4x\sqrt{-g} f(R) +S_m,
\label{act1}
\ee
 and corresponding field equation of metric to the action (\ref{act1}) is
\be
f^{\prime}(R)R_{\mu\nu}-\frac{1}{2}f(R)g_{\mu\nu}-\nabla_{\mu}\nabla_{\nu}f^{\prime}(R)+g_{\mu\nu}\Box f^{\prime}(R)=8\pi G T_{\mu\nu},
\ee
where $T_{\mu\nu} = -\frac{2}{\sqrt{-g}}\frac{\delta L_m}{\delta g^{\mu\nu}}$ is the energy momentum tensor. One can consider the energy-momentum of Maxwell field. However,   to show the mechanism in this section, we assume no matter ($T_{\mu\nu}=0$) and covariantly constant Ricci tensor, i.e., Ricci tensor is 
proportional to the metric $g_{\mu\nu}$. The field equation in this case reduces to 
\be
 f(R)-\frac{1}{2} Rf^{\prime}(R)=0.
\label{act3}
\ee
Eq. (\ref{act3}) provides a solution,
\be
A(r)= 1- \frac{R_0 r^2}{12} -\frac{M}{r},
\ee
for the space-time given by,
\be
ds^2 = -A(r)dt^2  + \frac{1}{A(r)}dr^2 + r^2 d\Omega^2,
\ee
where $R_0$ is the constant Ricci scalar and $M$ is the mass of the black hole. This space-time can be written similar as  Nariai space-time,
\be
ds^2=\frac{1}{\Lambda^2}\frac{1}{\cosh^2 x}(d\tau^2-dx^2)+ \frac{1}{\Lambda'^2} d\Omega^2,
\label{nar1}
\ee
where we defined  new coordinates $\tau$ and $x $ related to $t$ and $r$ via 
$t=2 r_0^2 \tau/[(1-R_0r_0^2/2) \epsilon]$ and $r=r_0+\frac{\epsilon}{2}(1+\tanh x)$ with $\Lambda^2=\frac{1 -r_0^2 R_0/2}{r_0^2}  ,\ \Lambda'=\frac{1}{r_0}$ ($\Lambda^2$ can be positive and negative). Here in Nariai space-time, two horizons  are separated by a small distance $\epsilon\rightarrow 0$($r_1=r_0+\epsilon$) at $r_0$ and $r_1$. $\Lambda$ becomes $1/r_{0}$ for $R_0=0$. To understand the behavior of horizon
we consider a more general space-time in terms of perturbations as follows,
\be
ds^2=\frac{e^{2\rho(x,\tau)}}{\Lambda^2}(d\tau^2-dx^2)+ \frac{ e^{-2\phi(x,\tau)}}{\Lambda'^2}d\Omega^2,
\label{nar11}
\ee
where $\rho(x,\tau)$ and $\phi(x,\tau)$ are given by
\ba
\rho=-\ln(\cosh x)+\delta\rho, 
\ea
\ba
\phi= \delta\,\phi.
\ea
We perturb the modified Einstein equations up to the first order. By considering  $F''(R_0)\neq 0$, these first order equations offer   
 a  solution which is given by
\be
\delta \rho=\rho_0\,\cosh \omega \tau\,\cosh^\beta x,\qquad\qquad \delta \phi= \phi_0\,\cosh \omega \tau\,\cosh^\beta x,
\ee
where $\omega$ and $\beta$ are constants determined by the field equations. The horizon radius $r_h$  is defined by
\be
g^{\mu\nu}\nabla_{\mu}\phi\nabla_{\nu}\phi=0. \label{horzcond}
\ee
From the solution $\delta \phi= \phi_0\,\cosh \omega \tau \,\cosh^\beta x$, we obtain
$\tanh^2\omega \tau=\tanh^2x$ because of $\omega^2=\beta^2$ \cite{Nojiri13}. It further simplifies the perturbation as  $\delta\phi=\delta\phi_h=\phi_0 (\cosh\beta \tau)^{\beta+1}$. From Eq. (\ref{nar11}), one now can map  $e^{-\phi}/\Lambda$ as  a radius coordinate and hence one may define the  dynamical horizon radius by using  Eq. (\ref{nar1}) as  
\be
r_h=\frac{e^{-\delta\phi_h}}{\Lambda'}= \frac{e^{-\phi_0\, (\cosh\beta \tau)^{\beta+1}}}{\Lambda'}. \label{rheqn}
\ee
The horizon radius in Eq. (\ref{rheqn}) can be increasing, decreasing or oscillatory depending on the parameters $\beta$ or $\omega$ and $\phi_0$.  For real positive values of $\beta+1$, we have increasing or decreasing horizon for $\phi_0$  negative or positive respectively. For  $\phi_0< 0$, the anti-evaporation occurs for $\beta+1$ positive. However, instabilty occurs in this case of anti-evaporation. For other case $\phi_0> 0$ with $\beta+1$ negative, we obtain stable anti-evaporation. For $\beta$ imaginary, we can get oscillatory solution.  Using this formalism, we study the anti-evaporation problem in $n$-dimensional space-time. It is possible to include quantum corrections, however, we will consider only the classical phenomenon in this paper. In  Sec. (\ref{sec:level3}), we consider Maxwell field and obtain a solution in $n$-dimension. The analytical solution may not be obtained as long as we consider non-conformal invariant action.  In order to get  analytical solutions for perturbations,  we require an analytical form of background solution. Therefore, we consider conformal invariant action for Maxwell field \cite{Sheykhi:2012zz}. 
\vspace{0.5cm}

\section{Field Equations in $N$-dimensional space-time}
\label{sec:level3}
As long as we assume conformally symmetric action for Maxwell field,  the analytical solution can be obtained. We choose non-linear form of action corresponding to Maxwell field to achieve such possibility. In this section we  present the field equations and discuss the solution for $f(R)$-gravity with conformally invariant Maxwell field
 in $n$-dimensional space time. We consider following action,
 
\be
S=\int d^nx \sqrt{-g}\left[f(R) - \left(F_{\mu\nu} F^{\mu\nu}\right)^p\right]. \label{action}
\end{equation}
Here $R$ is Ricci scalar,   $F_{\mu\nu}$  is electromagnetic field tensor and $p$ is a positive integer.  By varying the action  with respect to the metric $g_{\mu\nu}$ and  the
Maxwell field $A_{\mu}$ respectively, we obtain
\ba
  f'(R)R_{\mu\nu}-\frac{1}{2}f(R)g_{\mu\nu}+ g_{\mu\nu}\nabla^{\rho}\nabla_{\rho}f'(R)-\nabla_{\mu}\nabla_{\nu}f'(R)= \frac{T_{\mu\nu}}{2},\label{modeqn}
\ea
\hspace{0.5cm}
\ba
\nabla_{\mu}\left(F^{p-1}F^{\mu\nu}\right)=0, 
\label{eom}
\ea 

where, $f'(R)$ is the derivative of $f(R)$ with respect to $R$ and the energy momentum tensor may be written as,
\be
T_{\mu\nu}=4\left(p F^{p-1}F_{\mu\lambda}F^{\lambda}_{\nu}-\frac{1}{4} g_{\mu\nu}(F)^p\right).
\label{em}
\ee
We seek for a constant curvature solution, i.e., $R=R_0=$ constant. For such a case, trace of energy-momentum tensor should be zero. Under this condition,
we find $n=4p$. In addition, from Eq. (\ref{modeqn}) we also have ,
\be
R_0f'(R_0) -\frac{n}{2} f(R_0)=0. \label{eqnfR0}
\ee
Eq. (\ref{eqnfR0}) simplifies Eq. (\ref{modeqn}) as,
\be
f'(R_0)\left(R_{\mu\nu} - \frac{g_{\mu\nu}}{n}R_0\right)= \frac{T_{\mu\nu}}{2}
\ee
In $n$-dimensional space-time,  we consider the following line element
\be
ds^2= -N{(r)}dt^2+ \frac{1}{N(r)}dr^2 + r^2 d\Omega_{n-2}^2,
\ee
where, $d\Omega_{n-2}^2$ is the metric of an unit ($n$-$2$)-sphere and $N(r)$ is a static spherically symmetric function. In $n$-dimensional space-time
Ricci scalar turns out to be,
\be
R= - N''(r) -\frac{2(n-2)N'(r)}{r}+ \frac{(n-2)(n-3)}{r^2}-\frac{(n-2)(n-3)N(r)}{r^2} = R_0. \label{curv_eqn}
\ee
The solution for $N(r)$ corresponding to  Eq. (\ref{curv_eqn}) can be written as
\be
N(r)= 1 - \frac{2 m}{r^{n-3}} + \frac{C_1}{r^{n-2}} -C_2 r^2, \label{ndim_solution}
\ee
where, $C_1= \frac{q^2(-2 q^2)^{(n-4)/4}}{f'(R_0)}$,  $C_2= \frac{R_0}{n(n-1)}$, $m$ and $q$ are constants associated to the mass and
the charge of black hole respectively. It is noted that in this framework, the electric field behaves as in its standard form \cite{Sheykhi:2012zz}. The electric field in this case is given by $E=\frac{q}{r^{\frac{n-2}{2p-1}}}$ and takes its standard form for $n= 4p$. In the following subsection, we obtain the conditions for degenerate horizon.

\subsection{Extreme Black hole}
To investigate the instabilities and the evaporation/anti-evaporation, we consider a space-time near the degenerate horizon. In general, the black hole  has $n$ horizons in this theory. Depending on the values of parameters, $f(R)$ black hole may have degenerate horizons where two or more horizons
coincide. For such degenerate horizon, we have $N(r_0)=N'(r_0)=0$ which provides the following equations,
\be
N(r_0)= \frac{r_0^{n-2}-2 m r_{0} + C_1 - C_2 r_0^n}{r_0^{n-2}}=0,
\ee
or,
\be
r_0^{n-2}-2 m r_{0} + C_1 - C_2 r_0^n=0, \label{hor_eqn1}
\ee
and
\be
N'(r_0)=- 2 m (3-n) r_0^{2-n} + C_1(2-n)r_0^{1-n} - 2 C_2 r_0=0,
\ee
or
\be
- 2 (3-n) m r_0 + (2-n)C_1 - 2C_2r_0^n=0. \label{hor_eqn2}
\ee
We choose the value of $C_2$ from Eq. (\ref{hor_eqn1}),
\begin{eqnarray}
 C_2 = \frac{r_0^{(n-2)}- 2 m r_0 +C_1}{r_0^n}, \label{value_C2}
\end{eqnarray}
and we use this in  Eq. (\ref{hor_eqn2}) to obtain the value of $m$,
\begin{eqnarray}
 m = \frac{r_0^{(n-2)} + nC_1/2}{(n-1)r_0}. \label{value_m}
\end{eqnarray}
Substituting  Eq. (\ref{value_m}) in Eq. (\ref{value_C2}), we obtain,
\begin{eqnarray}
 C_2 =  \frac{1}{(n-1)} \Big[\frac{n-3}{r_0^2} - \frac{C_1}{r_0^n}\Big],
\end{eqnarray}
and which leads to
\begin{eqnarray}
 R_0 = n \Big[\frac{n-3}{r_0^2} - \frac{C_1}{r_0^n}\Big].
\end{eqnarray}


To define a nearly extreme black hole, we transform $r$  and $t$ in terms of $x$ and $\tau$ as,
\ba
r &=& r_{0} + \epsilon \cos(x), \\
t &=& \frac{2 \tau}{\epsilon N''(r_{0})},
\ea
where $\epsilon$ is very small. 
 A nearly extreme black hole will have the following form of function $N(r)$ \cite{Matyjasek:2013dua,Fernando:2016ksb},
\be
N(r)\approx \frac{N''(r_{0})}{2}(r-r_c)(r-r_h), \label{frnear_deg}
\ee
where, $r_c= r_{0} + \epsilon$ with $x=0$, $r_h=r_{0} - \epsilon$ with $x= \pi$. For such extreme black hole, we can write the metric in following
form,
\be
ds^2=\frac{2}{N''(r_{0})}\left(\sin^2 x d\tau^2 - dx^2 \right)+r^2_{0}d\Omega_{n-2}^2.
\label{nar2}
\ee

\begin{figure}[h]
\begin{tabular}{c c}
\includegraphics[width=0.5\linewidth]{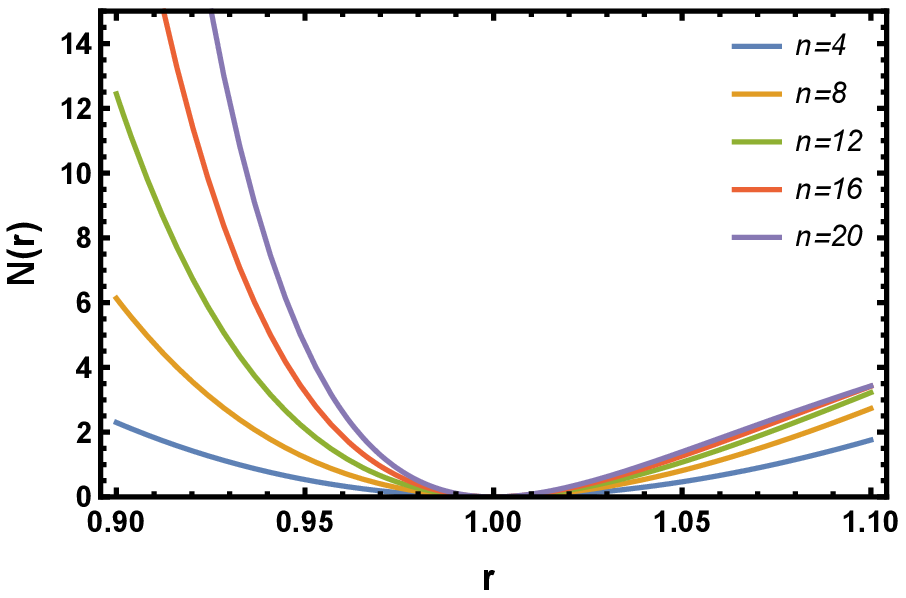}
  \includegraphics[width=0.5\linewidth]{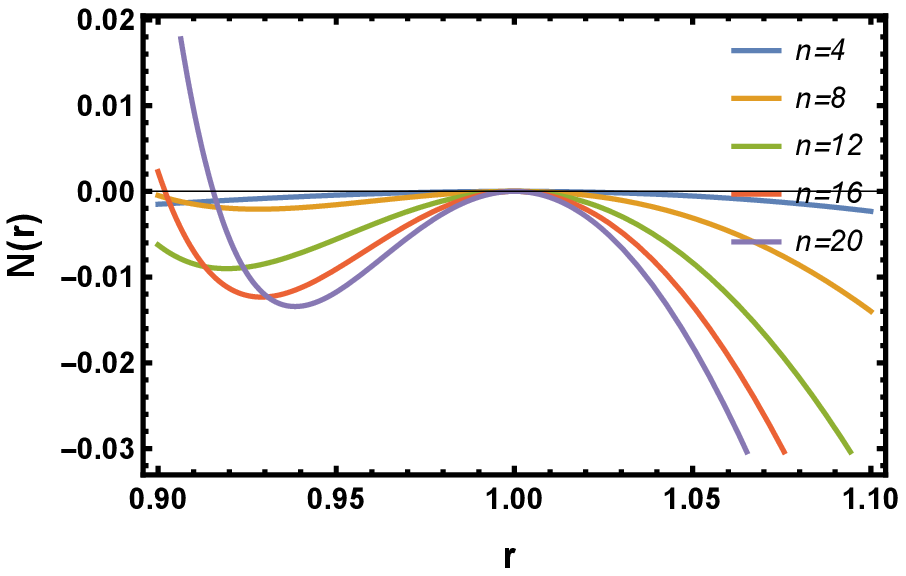}
\end{tabular}
\caption{\label{frplot} Left and right plots are for $N(r)$ with respect to $r$ for positive and negative $N''(r_{0})$  respectively with a degenerate horizon at $r_0=1$. }
\end{figure}
In Fig. (\ref{frplot}), we plot $N(r)$ where we set parameters such that we get the degenerate horizon $N(r_{0})=N'(r_{0})=0$. In  
   Sec. (\ref{sec:level4}), we study the perturbations around the solution for the extreme black hole. This two cases with positive and negative $N''(r_{0})$ will be considered while  discussing the solution of perturbations in  Sec. (\ref{sec:level4}). 
\section{\label{sec:level4}Anti-Evaporation/Evaporation}
In Fig. (\ref{frplot}), the plots for $N(r)$  are given  for  different values of $N''(r_0)$ and dimension $n$ with a degenerate horizon at $r_0=1$. The analytical definition of $N(r)$   near the degenerate horizon is given by Eq. (\ref{frnear_deg}). Near the degenerate horizon, the space-time is written as in Eq. (\ref{nar2}) in $n$-dimensions in different coordinate system.  In this section, we study the   anti-evaporation of $n$-dimensional black hole near the degenerate horizon. Here we implement perturbation analysis up to the linear order. It will be shown that  $f(R)$ theory is relevant for the dynamical nature of the perturbation $\delta \phi$. Setting $f''(R_0)=0$, one obtains $\delta \phi =0$, which indicates the constant horizon. Working with $f(R)$, the perturbation $\delta \phi$ becomes proportional to $\delta R$ and it gives a possibility of dynamical $\delta \phi$ . In simplifying constants we use the background equations. Maxwell field tensor  is defined by $\nabla_{\mu}(F^{p-1}F^{\mu\nu})=0$ and is modified in $n$-dimensions. 
\vspace{-0.5cm}
\subsection{\label{sec:level4.1}Perturbation}
 To know the behavior of the horizon, we introduce the fields $\rho(\tau,x)$ and $\phi(\tau,x)$ in  the space-time around the extreme black hole solution as follows,
 \begin{eqnarray}
  ds^2 =   \left(  \sin^2x \,d\tau^2 - d x^2\right) \frac{ e^{2  \rho(\tau,x)}}{\Lambda^2} 
  +  \frac{e^{-2 \phi(\tau,x)}}{\Lambda'^2} d\Omega_{n-2}^2,
 \end{eqnarray}
 where, $ \frac{1}{\Lambda^2} = \frac{2}{N''(r_0)} $ and $\frac{1}{\Lambda'^2}= r_{0}^2$.
 With this metric, the term $\nabla^{\rho}\nabla_{\rho}f'(R)$ in Eq. (\ref{modeqn}) turns out to be,
 \be
 \nabla^{\rho}\nabla_{\rho}f'(R) = e^{-2 \rho} \Lambda^2 \Big[\frac{1}{\sin^2{x}} \frac{\partial^2 f'(R)}{\partial \tau^2}-
 \cot{x} \frac{\partial f'(R)}{\partial x}-\frac{\partial^2 f'(R)}{\partial x ^2} 
 + (n-2) \left(- \frac{\dot{\phi}}{\sin^2 x} \frac{\partial f'(R)}{\partial \tau} + \phi'  \frac{\partial f'(R)}{\partial x}\right)\Big]
 \ee
 and components of Ricci tensor take the forms as,
 \ba
 R_{\tau \tau} &=& - \ddot{\rho} + \rho'' \sin^2 x + \sin{x} \cos{x} \rho' - \sin^2 x + (n-2) (\ddot{\phi}- \dot{\phi}^2)
 - (n-2) \dot{\phi} \dot{\rho} \nonumber \\ && - (n-2) \sin{x}\cos{x}\phi' - (n-2)\sin^2 x \rho' \phi', \label{rtautau} \\
 R_{x x} &=& \frac{\ddot{\rho}}{\sin^2 x} - \rho'' + 1 + (n-2) (\phi'' -\phi'^2) - (n-2)\left(\frac{\dot{\rho}\dot{\phi}}{\sin^2 x} + \phi' \rho'
 \right) -\rho' \cot{x}, \\
 R_{\tau x} &=& (n-2) \Big[ \dot{\phi}' - \dot{\phi} \rho' -\phi' \dot{\rho} - \dot{\phi}\phi'- \dot{\phi}\cot x \Big], \\
 R_{\theta_1 \theta_1} &=& \frac{\Lambda^2}{\Lambda'^2} e^{-2 (\rho+\phi)} \Big[\frac{1}{\sin^2 x} \left(\ddot{\phi} 
 + (2-n) \dot{\phi}^2\right) - \phi'' -\phi' \cot{x} + (n-2)\phi'^2\Big] + (n-3), \\
 R_{\theta_2 \theta_2} &=& \sin^2 \theta_1 R_{\theta_1 \theta_1} \\
   R_{\theta_3 \theta_3}&=& \sin^2 \theta_1 \sin^2 \theta_2 R_{\theta_1 \theta_1}....\mbox{so on.}                        
 \ea
 Ricci scalar in $n$-dimensional space-time may be written as,
 \ba
 R &=& \frac{\Lambda^2 e^{-2 \rho}}{\sin^2 x} \Big[- 2 \ddot{\rho} + 2 \rho'' \sin^2 x + 2(n-2)\ddot{\phi} -2 (n-2)\sin^2 x \phi''
 + 2 \rho' \sin x \cos x -2 (n-2) \phi' \sin x \cos x \nonumber \\ && - (n-1)(n-2)\dot{\phi}^2  + (n-1)(n-2)\sin^2 x \phi'^2 - 2 \sin^2 x
 \Big] + (n-2)(n-3)\Lambda'^2 e^{2\phi} \label{ricci}
 \ea
 Here,  primes and dots over $\phi$ or $\rho$ are derivatives  with respect to ``$x$'' and  ``$\tau$'',  respectively.
 In this space-time,  from  Eq. (\ref{eom}),  the electric field is given by,
 \begin{eqnarray}
  F_{x \tau}  = C q^\frac{1}{2p-1} e^{\frac{(n-2)\phi}{2p-1}}e^{2\rho}\sin x,
 \end{eqnarray}
where $C$ is a constant defined by $C\equiv \Lambda^{-2}[(-1)^p 2^{1-p} \Lambda'^{n-2}]^\frac{1}{2p-1}$, and $q$ is the charge of black hole. The  components of energy-momentum tensor can be written as
\begin{eqnarray}
 T_{\tau \tau} &=& 2 \left(p-\frac{1}{2}\right)\left(-2 q^2 \Lambda'^{2(n-2)} e^{2(n-2)\phi}\right)^\frac{p}{2p-1} \frac{\sin^2{x} e^{2 \rho}}{\Lambda^2}, \\ 
 T_{xx} &=& 2\left(p-\frac{1}{2}\right)\left(-2 q^2 \Lambda'^{2(n-2)} e^{2(n-2)\phi}\right)^\frac{p}{2p-1} \left(- \frac{e^{2 \rho}}{\Lambda^2}\right), \\ T_{x \tau} &=& 0, 
 \\ T_{\theta_1 \theta_1}&=& - \left(-2 q^2 \Lambda'^{2(n-2)} e^{2(n-2)\phi}\right)^\frac{p}{2p-1} \left( \frac{e^{-2 \phi}}{\Lambda'^2}\right),\\
 T_{\theta_2 \theta_2} &=& \sin^2 \theta_1 T_{\theta_1 \theta_1} \\
   T_{\theta_3 \theta_3}&=& \sin^2 \theta_1 \sin^2 \theta_2 T_{\theta_1 \theta_1}....\mbox{so on.}  
\end{eqnarray}
This leads the components of Eq. (\ref{modeqn}) to 
\begin{eqnarray}
 \frac{\Lambda^2}{\sin^2 {x}} e^{-2\rho} f'(R) \Big[-\ddot{\rho}+\rho'' \sin^2{x} + \sin{x}\cos{x} \rho' - \sin^2{x} 
 + (n-2)(\ddot{\phi} -\dot{\phi}^2 ) - (n-2) \dot{\phi}\dot{\rho} - (n-2)\sin{x}\cos{x} \phi' \nonumber \\
 - (n-2)\sin^2{x} \phi'\rho' \Big] -\frac{1}{2}f(R)  + e^{- 2 \rho}\Lambda^2 \Big[-\frac{\partial^2 f'(R)}{\partial x^2}
 + (n-2) \left(-\frac{\dot{\phi}}{\sin^2 x}\frac{\partial f'(R)}{\partial \tau} + \phi' \frac{\partial f'(R)}{\partial x}\right)\Big]\nonumber \\ + \frac{\Lambda^2}{\sin^2{x}} e^{- 2 \rho}  \Big[ \dot{\rho}\frac{\partial f'(R)}{\partial \tau} + \sin^2{x} \rho' \frac{\partial f'(R)}{\partial x} \Big]  =  \left(p-\frac{1}{2}\right)\left(-2 q^2 \Lambda'^{2(n-2)} e^{2(n-2)\phi}\right)^\frac{p}{2p-1}, \label{tteqn}
\end{eqnarray}

 \begin{eqnarray}
 -\frac{\Lambda^2}{\sin^2 {x}} e^{-2\rho} f'(R) \Big[\ddot{\rho}- \rho'' \sin^2{x}  + \sin^2{x} + (n-2)\sin^2{x} \left(\phi''
 -\phi'^2\right) - (n-2)\left(\dot{\phi}\dot{\rho} + \phi'\rho' \sin^2{x}\right) -\rho'\sin{x}\cos{x} \Big]
 \nonumber \\ -\frac{1}{2}f(R)  + e^{- 2 \rho}\Lambda^2 \Big[ \frac{1}{\sin^2{x}} \frac{\partial^2 f'(R)}{\partial \tau^2}
 - \cot{x} \frac{\partial f'(R)}{\partial x }+ (n-2) \left(-\frac{\dot{\phi}}{\sin^2 x}\frac{\partial f'(R)}{\partial \tau} + \phi' \frac{\partial f'(R)}{\partial x}\right)\Big]\nonumber \\ - \frac{\Lambda^2}{\sin^2{x}} e^{- 2 \rho}  \Big[ \dot{\rho}\frac{\partial f'(R)}{\partial \tau} + \sin^2{x} \rho' \frac{\partial f'(R)}{\partial x} \Big]  =  \left(p-\frac{1}{2}\right)\left(-2 q^2 \Lambda'^{2(n-2)} e^{2(n-2)\phi}\right)^\frac{p}{2p-1},\label{xxeqn}
\end{eqnarray}
\begin{eqnarray}
 (n-2)f'(R)\Big[\dot{\phi}' - \dot{\phi}\rho' -\phi'\dot{\rho} - \dot{\phi}\phi' - \dot{\phi}\cot{x}\Big]
 -\Big[\frac{\partial^2 f'(R)}{\partial \tau \partial x} - (\rho' + \cot{x})\frac{\partial f'(R)}{\partial \tau} -\dot{\rho}
 \frac{\partial f'(R)}{\partial x}\Big]=0,\label{txeqn}
\end{eqnarray}

 \begin{eqnarray}
 \frac{\Lambda^2}{\sin^2 {x}} e^{-2\rho} f'(R) \Big[ \ddot{\phi} + (2-n)\dot{\phi}^2 - \sin^2{x} \phi'' 
 - \phi' \sin{x}\cos{x} + (n-2)\sin^2{x}\phi'^2\Big] + (n-3)e^{2 \phi} \Lambda'^2 f'(R) -\frac{1}{2} f(R) \nonumber \\
 + \Lambda^2 e^{- 2\rho} \Big[\frac{1}{\sin^2{x}}\frac{\partial^2 f'(R)}{\partial \tau^2} - \cot{x} \frac{\partial f'(R)}{\partial x} - \frac{\partial^2 f'(R)}{\partial x^2} + (n-2)\left(-\frac{\dot{\phi}}{\sin^2{x}}\frac{\partial f'(R)}{\partial \tau} + 
 \phi' \frac{\partial f'(R)}{\partial x}\right)\Big] \nonumber \\ 
 - e^{-2 \rho} \Lambda^2 \left(-\frac{\dot{\phi}}{\sin^2{x}}\frac{\partial f'(R)}{\partial \tau} + 
 \phi' \frac{\partial f'(R)}{\partial x}\right)= -\frac{1}{2}\left(-2 q^2 \Lambda'^{2(n-2)} e^{2(n-2)\phi}\right)^\frac{p}{2p-1}. \label{ththeqn}
\end{eqnarray}
 We now perturb the fields $\phi=\delta \phi(\tau,x)$ and $\rho=\delta \rho(\tau,x)$  around Nariai-like space-time. We also perturb
 Ricci scalar $R$ around its constant background $R_0$.  From Eq. (\ref{txeqn}) we obtain, 
 \begin{eqnarray}
  \delta R= \frac{(n-2)f'(R_0)}{f''(R_0)}\delta \phi.
 \end{eqnarray}
Here, we note that the perturbation $\delta \phi$ vanishes if $f''(R_0)=0$,  indicating no possibility of evaporation or anti-evaporation even in $n$-dimension. Eq. (\ref{ththeqn})  provides a differential equation of the perturbation $\delta \phi(\tau,x)$,
\begin{eqnarray}
 (n-1)\Lambda^2 \Big[\frac{\ddot{\delta \phi}}{\sin^2{x}}  - \delta \phi'' - \cot{x} \delta \phi' \Big]  
 + \Big[ n(n-3) \Lambda'^2 - \frac{ (n-2)f'(R_0)}{2 f''(R_0)}  + \frac{ p (n-2)}{f'(2 p-1)} q' \Big]\delta \phi=0, \label{prteqn1}
\end{eqnarray}
and Eq.  (\ref{tteqn}) or (\ref{xxeqn})  can be written as,
\begin{eqnarray}
 -\frac{\ddot{\delta \rho}}{\sin^2{x}}  + \delta \rho'' + \cot{x} \delta \rho'  + 2 \delta \rho  +  (n-2) \Big[\frac{\ddot{\delta \phi}}{\sin^2{x}}  - \delta \phi'' - \cot{x} \delta \phi'\Big] 
 - (n-2) \Big[1 +  \frac{f'}{2 f''\Lambda^2}\nonumber \\ + \frac{p  q'}{f'(R_0) \Lambda^2}\Big] \delta \phi =0, \label{prteqn2}
\end{eqnarray}
where $q' =  [-2 q^2 \Lambda'^{2 (n-2)})]^\frac{p}{2p-1}$.

Under the coordinate transformation $dx=\sin{x} du$, Eq.(\ref{prteqn1}) becomes, 
\begin{eqnarray}
 \ddot{\delta \phi}- \delta  \phi_{, {uu} }  + \alpha \cosh^{-2} (u+c)\delta \phi = 0, \label{eqn_deltaphi}
\end{eqnarray}
where ``$_{,u}$'' denotes $\partial /\partial u$, c is an integral constant and $\alpha$ is a constant given by 
\begin{eqnarray}
\alpha=\frac{1} {(n-1)\Lambda^2}\Big[ n(n-3) \Lambda'^2 - \frac{ (n-2)f'(R_0)}{2 f''(R_0)}  + \frac{ p (n-2)}{f'(2p-1)} q' \Big].
\end{eqnarray}

A solution for Eq. (\ref{eqn_deltaphi}) is  $\delta \phi =  \left(A e^{\omega \tau} + B e^{-\omega \tau}\right) \cosh^{\beta} (u+c)$, with
\begin{eqnarray}
\omega^2 - \beta^2=0, \ \ \  0=\alpha+\beta(\beta -1),
\end{eqnarray}
which gives
\begin{eqnarray}
\omega = \pm \beta,\ \ \beta =\beta_{\pm}\equiv\frac{1}{2}(1\pm\sqrt{1-4\alpha}). \label{betaeq1}
\end{eqnarray}
Using the horizon condition from Eq.(\ref{horzcond}), we have
\begin{eqnarray}
\tanh{(u+c)}= \frac{A e^{\omega \tau} - B e^{-\omega \tau}}{A e^{\omega \tau} + B e^{-\omega \tau}}.
\end{eqnarray}

Then we find that 
\begin{eqnarray}
\delta \phi = \frac{\left(A e^{\beta \tau} + B e^{-\beta \tau}\right)^{\beta +1}}{(2\sqrt{A B})^{\beta}},
\end{eqnarray}
To know the behaviour of $\delta \phi$, we consider a case where $A=B$, this leads the expression of $\delta \phi$ as,
\begin{eqnarray}
 \delta \phi = 2 A [\cosh{(\beta \tau)}]^{\beta +1},
\end{eqnarray}
and according to Eq.(\ref{rheqn}), the radius of the horizon is given by 
\begin{eqnarray}
r_h = \frac{e^{- 2A [\cosh{(\beta \tau)}]^{\beta +1}}}{\Lambda'}.
\end{eqnarray}
The negative value of $\beta+1$ can be obtained with  negative root of $\beta$. It is noted that if $\beta +1$ is negative which we can see in case (a.) discussed below (where $\Lambda^2$ is negative) by setting, e.g., $n=8$ and $n'=-1$, the perturbation $\delta \phi$ decreases and thus horizon size increases for positive values of $A$ and which is the case of anti-evaporation. After a certain time, the horizon size becomes a constant, $r_h=\frac{1}{\Lambda'}$. For positive $\beta + 1$, e.g. $n=4$ and $n'=-1$ with positive root in the case (a.), evaporation occurs. The positive value of $\beta+1$ can also explain anti-evaporation with  negative $A$. However, in this  case of positive $\beta + 1$ instability occurs.   We can also have solution for $\delta \rho$ in terms of $\delta \phi$, given by $\delta \rho = \gamma \delta \phi$  satisfying  Eqs. (\ref{prteqn1}) and (\ref{prteqn2}), where $\gamma = \frac{(n-2)(C_3 + \alpha)}{(2 +\alpha)}$ and $C_3$ is given by
\begin{eqnarray}
 C_3 = 1 +  \frac{f'}{2 f''\Lambda^2}  + \frac{p  q'}{f'(R_0) \Lambda^2}.
\end{eqnarray}
 The perturbation $\delta \rho$ evolves in the same way as $\delta\phi$. One can eleminate $q'$ from the background equation. From the background Einstein equation, one can find,
\begin{eqnarray}
 p q' = - f'(R_0)\left(\Lambda^2 + (n-3) \Lambda'^2\right),
\end{eqnarray}
which results
\begin{eqnarray}
 \alpha = -\frac{2}{n-1} - \frac{(n-2) f'(R_0)}{2(n-1)f''(R_0)\Lambda^2} +  \frac{(n-2)(n-3)\Lambda'^2}{(n-1)\Lambda^2}, 
\end{eqnarray}
where, we used $n=4p$. Here the constants can be computed at an extreme horizon,
\begin{eqnarray}
 \Lambda^2 &=& -\frac{(n-3)}{r_0^2} + \frac{C_1 n}{2 r_0^n}, \\
 \Lambda'^2 &=& \frac{1}{r_0^2}.
\end{eqnarray}
Now we consider two cases (a.)  and (b.). In the first case, we assume the condition $\frac{(n-3)}{r_0^2} \gg \frac{C_1 n}{2 r_0^n}$ and in the second case we consider the opposite way.
\subsubsection{Case (a.)}
In this case, we have $\Lambda^2\approx - \frac{(n-3)}{r_0^2}$.  If one considers a theory $f(R)\sim  R^{n'}$, then  $n'$ and the dimension of space-time $n$ can determine the value of constant $\alpha$ as following,
\begin{eqnarray}
 \alpha = -\frac{n}{(n-1)} + \frac{n(n-2)}{2(n-1)(n'-1)},
\end{eqnarray}
and the term $1- 4 \alpha$ in Eq. (\ref{betaeq1}) can be positive if,
\begin{eqnarray}
 n' \geq  \frac{(2n-1)(n+1)}{(5n -1)} \ \ \mbox{and} \ \ n' \leq  1 .
\end{eqnarray}
For $n=4$, we get $n' \geq 1.84$ for real value of $\beta$. We can get positive and negative values of $1+\beta$ with different choices of $n$ and $n'$. For example, as already mentioned, $n' =-1$
can explain both anti-evaporating and  evaporation for negative and positive values of $\beta+1$ with positive $A$.  For imaginary values of $\beta$, we have oscillatory solution which we will discuss bellow. 
\subsubsection{Case (b.)}
\begin{figure}[h]
\begin{tabular}{c c}
\hspace{-0.2cm}\includegraphics[width=0.45\linewidth]{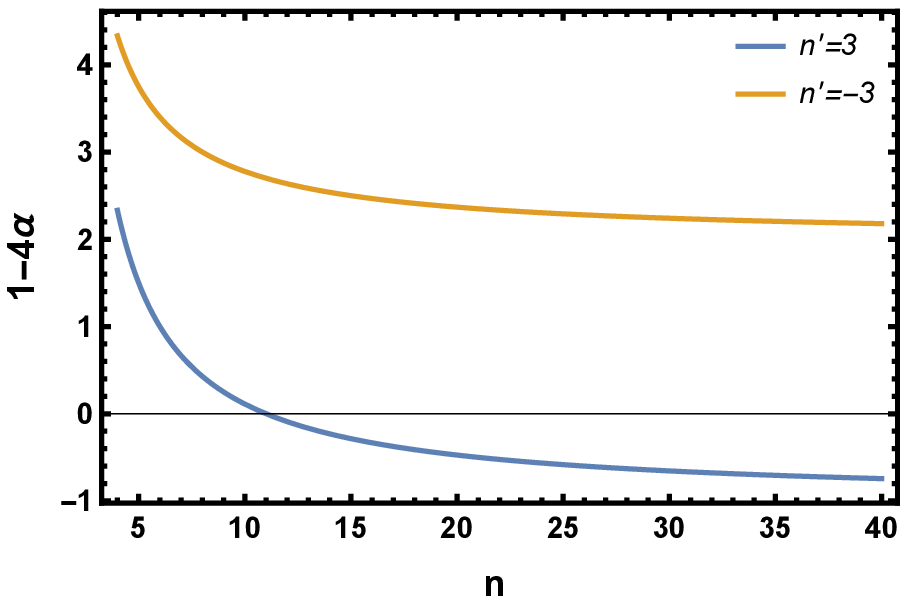}
 \includegraphics[width=0.453\linewidth]{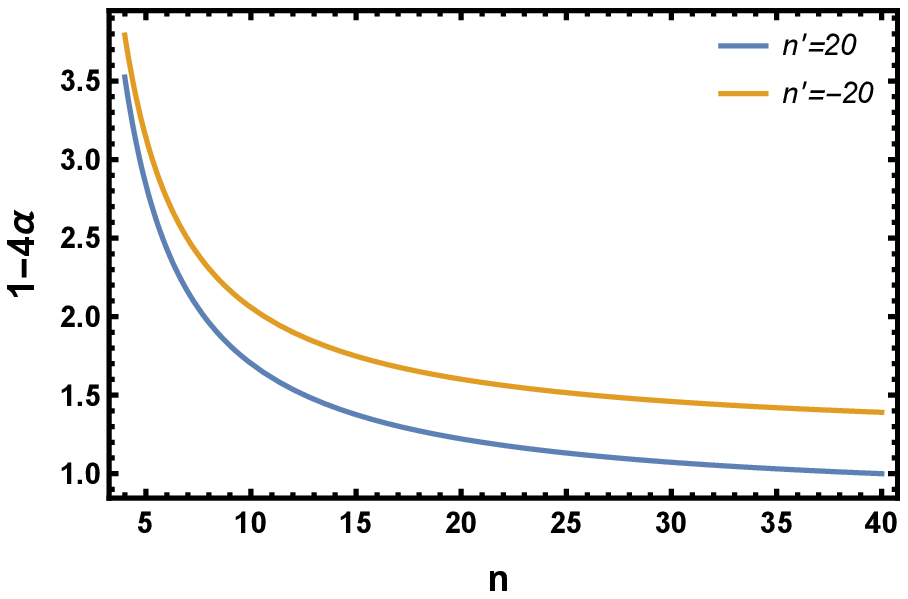}
\end{tabular}
\caption{\label{fig:alpha} The plot of $1- 4 \alpha$ with dimension $n$ of space-time for $r_0= 0.1$ and $C_1 =10$. }
\end{figure}

In this case where  $\frac{C_1}{2 r_0^n} \gg \frac{(n-3)}{r_0^2}$, under similar theories, we find,
\begin{eqnarray}
 \alpha = -\frac{2}{(n-1)} +  \frac{(n-2)}{(n-1)} \Big[\frac{1}{ (n'-1)} +\frac{2(n-3)r_0^{(n-2)}}{n C_1}\Big].
\end{eqnarray}
 For $n=4$ and $n'=2$, we obtain very small positive value of $\alpha$ resulting only  decreasing horizon for positive and negative roots with  positive $A$. We plot $1- 4 \alpha$ with respect to dimension $n$ in the left panel in Fig. (\ref{fig:alpha}) for $n'=\pm3$. It is observed on the left  panel of Fig. (\ref{fig:alpha}) that initially for lower $n$, the first negative term is dominant and becomes smaller after a certain value of $n$ making the whole term $1- 4 \alpha$ nearly constant for both positive and negative value of $n'$. However, for negative value of $n'$, the term $1- 4 \alpha$ remains positive for all large values of $n$.  For large value $n'$,  both curves approach to each other for both positive and negative values and converge to nearly unity since $\alpha$ becomes very small as shown in the right panel of Fig. (\ref{fig:alpha}).

For negative value of $1 - 4 \alpha$, i.e., imaginary value of $\beta$, we observe oscillatory solution. Let us consider $\sqrt{1- 4 \alpha} =  i y$, where $y$ takes positive and negative values. We can write the real solution of $\delta \phi$ as,
\begin{eqnarray}
 \delta \phi = 2 A e^{\frac{3 \gamma}{2}  - \frac{y \theta}{2}} \cos{\left(\frac{y \gamma}{2} + \frac{3 \theta}{2}\right)} , \label{osci_sol}
\end{eqnarray}
where,
\begin{eqnarray}
 \gamma = \ln{\sqrt{\cos^2{\frac{y \tau}{2}}\cosh^2{\frac{ \tau}{2}}+\sin^2{\frac{y \tau}{2}}\sinh^2{\frac{ \tau}{2}}}},
\end{eqnarray}

and $\theta$ is given by
\begin{eqnarray}
 \theta = \tan^{-1}{\Big[\tan{\frac{y \tau}{2}} \tan{\frac{\tau}{2}   }\Big]} + \theta_1,
\end{eqnarray}
where $\theta_1$ is the phase term given by
\begin{eqnarray}
 \theta_1 = \pi, \ \mbox{for} \ \  \cos{\frac{y \tau}{2}}\cosh{\frac{ \tau}{2}}<0 \ \  \mbox{and} \ \sin{\frac{y \tau}{2}}\sinh{\frac{ \tau}{2}} \geq 0, \nonumber \\
 \theta_1 = - \pi, \ \mbox{for} \ \  \cos{\frac{y \tau}{2}}\cosh{\frac{ \tau}{2}}<0 \ \  \mbox{and} \ \sin{\frac{y \tau}{2}}\sinh{\frac{ \tau}{2}} < 0,  \nonumber  \\
 \theta_1 = 0 \ \ \mbox{for} \cos{\frac{y \tau}{2}}\cosh{\frac{ \tau}{2}} >0.
\end{eqnarray}
However, we can remove the phase term $\theta_1$ as it is a constant term in the solution of $\delta \phi$. The solution given in Eq. (\ref{osci_sol}) is oscillatory. The amplitude of the oscillation increases exponentially in this case, thus exhibiting instability.  We can see that the given solution is independent of the sign of $y$ and $\tau$. The same solution can be obtained with the negative root, i.e. with $-y$.  We note that the instability can not be controlled  by parameter $\beta$ (or $y$). \\

We have discussed above non-oscillatory and oscillatory solutions. Both solutions can explain anti-evaporation and evaporation. However, non-oscillatory solutions are stable and unstable, on the other hand, oscillatory solutions are only unstable. In non-oscillatory case, the phenomenons of anti-evaporation/evaporation can survive for long time with some specific values of parameters, e.g. for $\beta +1$ very small.   


\section{conclusion}\label{sec:level5}
General Relativity predicts constant horizon around the Nariai-like space-time for Reissner-Nordstr\"{o}m black hole during its evolution at classical level. In contrast, $f(R)$-gravity offers a possibility of dynamical behaviour of the degenerate horizon in this black hole which could even be possible in Schwarzschild black hole \cite{Nojiri:2014jqa}. In this work, we generalize the theory in $n$-dimension to broaden the implications of $f(R)$-gravity. First it was found  that General Relativity with $n$-dimension still does not explain anti-evaporation and evaporation. This can be realized if one sets $f''(R_0)=0$ in perturbation equations. Therefore, despite $n$-dimension is richer, it does not help in anti-evaporation and evaporation unless we replace General Relativity by 
$f(R)$-gravity. Considering $f(R)$-gravity, we obtained the dynamical equation for the degenerate horizon and we categorize solutions in three types. One in which, we obtained increasing solution for horizon with positive constant $A$ and negative $\beta+1$, which indicates anti-evaporation and is stable. After a certain time the horizon size becomes a constant. Anti-evaporation also occurs with the negative value of $A$ with positive $\beta+1$, however, instability is associated with anti-evaporation in this case.  Second, where evaporation occurs, is decreasing solution for horizon. The stable evaporation can be explained with negative  $A$ and $\beta+1$. Likewise to anti-evaporation, evaporation can also occur for positive $A$ and positive $\beta+1$ with instability. The last one is oscillating solution with increasing amplitude. Both the degenerate horizon and Ricci scalar oscillate in the same way, since the perturbation of the degenerate horizon is proportional the perturbation field of Ricci scalar. In this paper, it is noted that a black hole can have anti-evaporation at classical level and this effect can remain for a long time.   This offers a possibility of long-lived primordial black hole even with smaller mass.

%

\end{document}